\documentclass[aps,prl,reprint]{revtex4-1}

\usepackage{amsmath,amsfonts,amssymb,graphicx,braket,bbold,mathtools,array}
\usepackage{float}
\usepackage{multirow}
\usepackage[]{units}
\usepackage[T1]{fontenc}
\usepackage{xr-hyper}
\usepackage{hyperref}
\hypersetup{
    colorlinks=true,
    linkcolor=blue,
    filecolor=blue,      
    urlcolor=blue,
    citecolor=blue
}

\newcommand*{\figref}[2][]{%
  \hyperref[{fig:#2}]{%
   \ref*{fig:#2}%
    \ifx\\#1\\%
    \else #1%
    \fi
  }%
}

\begin{document}
\title{Cavity Control over Heavy-Hole Spin Qubits in Inversion-Symmetric Crystals}
\author{Philipp M. Mutter}
\email{philipp.mutter@uni-konstanz.de}
\author{Guido Burkard}
\email{guido.burkard@uni-konstanz.de}
\affiliation{Department of Physics, University of Konstanz, D-78457 Konstanz, Germany}

\begin{abstract}
The pseudospin of heavy-holes (HHs) confined in a semiconductor quantum dot (QD) represents a promising candidate for a fast and robust qubit. While hole spin manipulation by a classical electric field utilizing the Dresselhaus spin-orbit interaction (SOI) has been demonstrated, our work explores cavity-based qubit manipulation and coupling schemes for inversion-symmetric crystals forming a planar HH QD. Choosing the exemplary material Germanium (Ge), we derive an effective cavity-mediated ground state spin coupling that harnesses the cubic Rashba SOI. In addition, we propose an optimal set of parameters which allows for Rabi frequencies in the MHz range, thus entering the strong coupling regime of cavity quantum electrodynamics.
\end{abstract}
\maketitle
After decades of research, a full-scale universal quantum computer is still not available. Various platforms are investigated in pursuit of this ultimate goal including ultra-cold atoms \cite{Saffman2016}, trapped ions \cite{Blatt2012,Bruzewicz2019}, superconducting circuits \cite{Chow2012, Barends2014} and highly engineered semiconductor structures \cite{Kloeffel_review2013, Zhang_review2019}. Recently, holes (missing electrons in the valence band) confined in QDs defined in the group-IV material Ge have attracted a fair amount of renewed interest due to state-of-the-art experiments~\cite{Watzinger2018,Vukusic2018, Hardy2019, Hendrickx2020, Hendrickx2020b, Hofmann2019arXiv,Froning2020arXiv,Wang2020arXiv}, placing them at the forefront of promising candidates for quantum information processing. There are two main points in favour of a qubit, the quantum analogue of a classical bit, built from the pseudo-spin degree of freedom of holes in Ge QDs. (i) They are reliable due to long coherence times owing to the absence of a valley degree of freedom, low disorder and weak isotropic hyperfine interaction (HFI), the latter being a consequence of the p-symmetry of the orbital wave functions. While there are other types of HFI in hole systems~\cite{Fischer2009}, these can be suppressed by working with isotopically pure samples. (ii) Qubit manipulation is potentially fast and may be performed by all-electrical means. This is made possible by the strong intrinsic SOI of holes in Ge, which can be used in an electric dipole spin resonance (EDSR) scheme to manipulate spins in the context of spintronics \cite{Hao2010}. EDSR by an alternating electric field in electron systems has been found to enable fast single-qubit operations~\cite{Golovach2006, Li2013, Liu2018}. Theoretical investigations of HHs in semiconductor structures include the detailed study of 1D-materials such as nano- and hutwires, investigating decoherence and relaxation times as well as EDSR~\cite{Kloeffel2011,  Kloeffel2013a, Kloeffel2013b, Watzinger2016, Kloeffel2018}.

Previous studies on HHs in planar QDs have identified the Dresselhaus SOI as the primary driving force of HH spin manipulation by a classical electric field~\cite{Bulaev2005, Bulaev2005b, Bulaev2007} as it couples the ground state to the first excited state with opposite spin. Holes in bulk inversion-symmetric crystals such as Ge, however, are not subject to this type of SOI~\cite{Terrazos2018arXiv}, putting at risk fast qubit manipulation. In this paper we show that with the aid of the  sizeable intrinsic Rashba SOI of holes and an externally applied in-plane magnetic field the ground state HH spins may still be rotated coherently by coupling the QD to a photonic microwave resonator in the framework of cavity quantum electrodynamics. Moreover, we propose a set of system parameters which allows for Rabi frequencies of several MHz. 

We consider a planar QD formed in a heterostructure with an inversion-symmetric middle layer, such as  Ge/Si$_y$Ge$_{1-y}$ quantum wells~\cite{Kuo2005, Paul2008, Dobbie2012}. As such we assume strong confinement along the growth direction (say $z$) of the heterostructure such that only the lowest orbital energy level is occupied along that direction. The strong confinement induces a splitting $\Delta = 2\hbar^2 \gamma_s/m_0 d_z^2$ between the HH and light hole (LH) bands, where $\gamma_s$ is the Luttinger parameter in spherical approximation, $m_0$ is the bare electron mass and $d_z$ is the height of the QD, i.e., the thickness of the middle layer in a heterostructure. Consequently, only the HH band is occupied at low temperatures, $k_B T \ll \Delta$, and we may adopt an effective description of the states with Kramers index (`spin') $ J_z \equiv s = \pm3/2$. For parabolic and circular in-plane confinement characterized by the energy scale $\hbar \omega_0$ and an out-of-plane magnetic field $B$ the effective Hamiltonian describing a HH confined in a planar QD reads
	\begin{align} \label{eq:H_0}
		H_0 = \frac{\pi_x^2 + \pi_y^2}{2m} + \frac{1}{2}m  \omega_0^2 (x^2 + y^2) + \frac{1}{2} g_z \mu_B B \sigma_z,
	\end{align}
where $\boldsymbol{\pi} = \mathbf{p} + e  \mathbf{A}$, and $\mathbf{A} = B(-y,x,0)/2$ denotes the vector potential in symmetric gauge, $m$ is the in-plane HH mass, $g_z >0$ is the out-of-plane HH g-factor, and $\sigma_z$ the Pauli matrix along the quantization axis. The eigenstates of this Hamiltonian are $\vert n,l,s \rangle \equiv \vert n,l \rangle  \vert s \rangle$, where $\vert n, l \rangle$ denote the Fock-Darwin states with principal (azimuthal) quantum number $n$ ($l$). The corresponding energies are given by $E_{n,l,s} = l \hbar \omega_L + (n+1) \hbar \omega + s g_z \mu_B B/3$, where $ \omega_L = eB/2m$ is the Larmor frequency and $\omega = \sqrt{\omega_0^2 + \omega_L^2}$~\cite{Fock1928}. Note that the Larmor frequency can be quite sizeable for HH in planar QDs due to its dependence on the inverse mass. 

The Hamiltonian in Eq.~\eqref{eq:H_SO} is separable in orbital and spin parts, i.e., it does not mix hole states of different pseudo-spin. Mixing between these states enters via the spin-orbit interaction, which for HHs in inversion-symmetric crystals is described by the Hamiltonian~\cite{Bulaev2005, Bulaev2007, Moriya2014, Miserev2017},
	\begin{align} \label{eq:H_SO}
		H_{\text{SO}} = i \alpha \left( \sigma_+ \pi_-^3 - \sigma_- \pi_+^3 \right) + \xi (b) \left(  \sigma_+ \pi_-^2 + \sigma_- \pi_+^2  \right),
	\end{align}
where $\sigma_{\pm} = (\sigma_x \pm i \sigma_y)/2$ with in-plane Pauli matrices $\sigma_{x/y}$ and $\pi_{\pm} = \pi_x \pm i \pi_y$. The first term in Eq.~\eqref{eq:H_SO} is obtained in second-order perturbation theory from the Luttinger-Kohn Hamiltonian including the Rashba term $\alpha_R \left(\boldsymbol{\pi} \times \langle E_z \rangle \mathbf{e}_z \right)\cdot \mathbf{J}$, where $\mathbf{J}$ is a vector containing spin-3/2 matrices, $\alpha_R$ is the Rashba coefficient and $\langle E_z \rangle$ is the averaged electric field along the growth direction experienced by the hole in an asymmetric quantum well. One finds $\alpha = 3 \gamma_s \alpha_R \langle E_z \rangle/2 m_0 \Delta$, and we define the quantity $\lambda_R= \alpha  \left( m \hbar \omega_0   \right)^{3/2}$ as the characteristic energy scale of the cubic Rashba SOI. The second term in Eq.~\eqref{eq:H_SO} is only present when an in-plane magnetic field is applied to the system and stems from the combined effects of the SOI and HH-LH mixing. Due to in-plane degeneracy, we may choose a coordinate system such that the additional in-plane field is along $x$, resulting in a total magnetic field $\mathbf{B} = (b, 0, B)$ and $\xi(b) = 3 \gamma_s \kappa  \mu_B b/m_0 \Delta$, where $\kappa$ is the magnetic Luttinger parameter in the context of the envelope function approximation. We define the energy scale of the magnetic coupling strength as $\lambda_b = \xi (b) m \hbar  \omega_0 $.

Note that applying an in-plane magnetic field also introduces an in-plane Zeeman term $H_{Z}^x = g_x \mu_B b \sigma_x/2$ to the effective HH Hamiltonian via the cubic spin-3/2 operator terms. The in-plane g-factor $g_x$ is typically much smaller than its out-of-plane counterpart, $g_x \ll g_z$, and we may treat $H_{Z}^x$ as a perturbation. On the other hand, we completely disregard the orbital effects of the in-plane magnetic field due to strong out-of-plane confinement (of the order of $10$ meV for a typical dot height of 10 nm). The latter is valid as long as $\hbar eb/2m \ll U_z = \hbar^2 \pi^2/2m d_z^2$, where $d_z$ is the height of the QD. We can express the condition in terms of the HH-LH splitting $\Delta$, $b \ll 10^4 \Delta[\text{eV}]$~T. 
	\begin{figure}
		\includegraphics[scale=0.7]{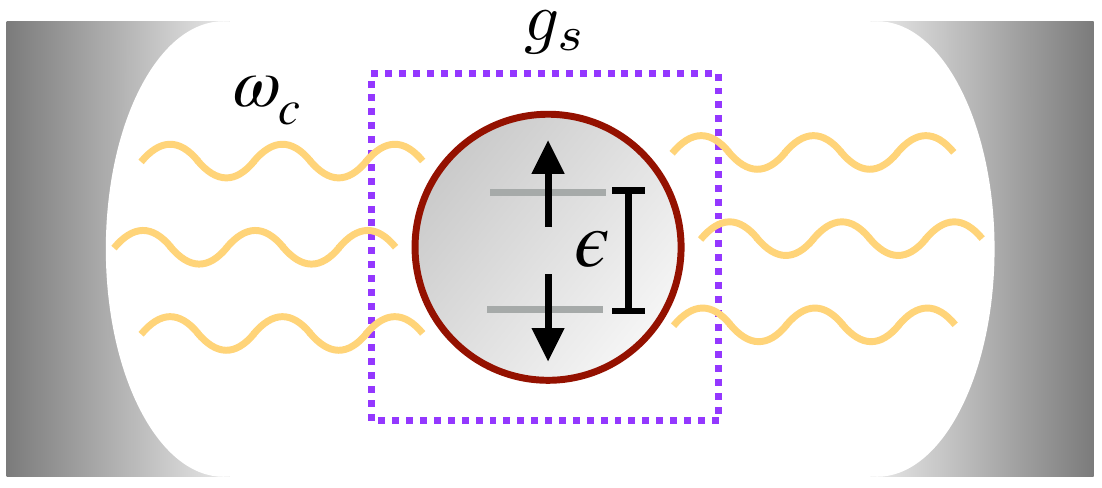}
		\caption{Coupling of a quantum dot to a resonator. Photons of energy $\hbar  \omega_c$ are confined in a cavity and interact with a planar quantum dot with renormalized ground state spin splitting $\epsilon$. The dotted square shows the interaction region, the strength of the effective spin-photon coupling being $g_s$.}
		\label{fig:cavity_QD_sketch}
	\end{figure}
	
As depicted in Fig.~\ref{fig:cavity_QD_sketch}, we aim to couple the ground state spins to a cavity mode of frequency $\omega_c$, described by the Hamiltonian $H_c = \hbar \omega_c a^{\dagger} a$, where $a^{\dagger}$ and $a$ are the photon creation and annihilation operators, respectively. The interaction of the cavity photons with the QD holes in the Fock-Darwin-spin basis $\lbrace \vert \nu \rangle \equiv \vert n ,l,s \rangle \rbrace$ is of the form
	\begin{align} \label{eq:cavity_coupling}
	\begin{split}
		&H_I = \hbar g_c  \left(a + a^{\dagger} \right)  \sum_{ \nu , \nu'}  \Pi^{\nu}_{ \nu'}  \vert \nu \rangle \langle  \nu' \vert,
	\end{split}
	\end{align}
where $g_c = e \sqrt{\omega_0/2 \epsilon_0 \epsilon_r m V \omega_c}$~\cite{Cohen-Tannoudji1989, Burkard2006} for a cavity of volume $V$ and relative permittivity $\epsilon_r$, and $\Pi^{\nu}_{\nu'}~\equiv~\frac{1}{\sqrt{\hbar \omega_0 m}}\langle n, l, s \vert  \pi_x \vert n',l',s' \rangle$ are the dimensionless momentum matrix elements for linearly polarized light along $x$, i.e., along the in-plane magnetic field.

The lowest energy state that the ground state is coupled to by the in-plane magnetic field term $\propto b$ in Eq.~(\ref{eq:H_SO}) is $\vert 2, \pm 2, \mp 3/2 \rangle$ and this state can transition via the combined effects of the Rashba SOI, the in-plane magnetic field and the electric dipole coupling to the orbital ground state with opposite spin, thereby creating an effective ground state spin-photon coupling. For the spin to flip we need an odd number of transitions due to the Rashba SOI and in-plane magnetic field. Since the ground state spins are not directly coupled, we expect the minimum number of spin-orbit induced transitions required to be three. A graphical overview of the allowed transitions in the low energy part of the system is given in Fig.~\figref[a]{single_QD_coupling_schematic}, and an exemplary sequence of transitions realizing a ground-state spin-flip is shown in Fig.~\figref[b]{single_QD_coupling_schematic}.
	\begin{figure*}
			\includegraphics[scale=0.5]{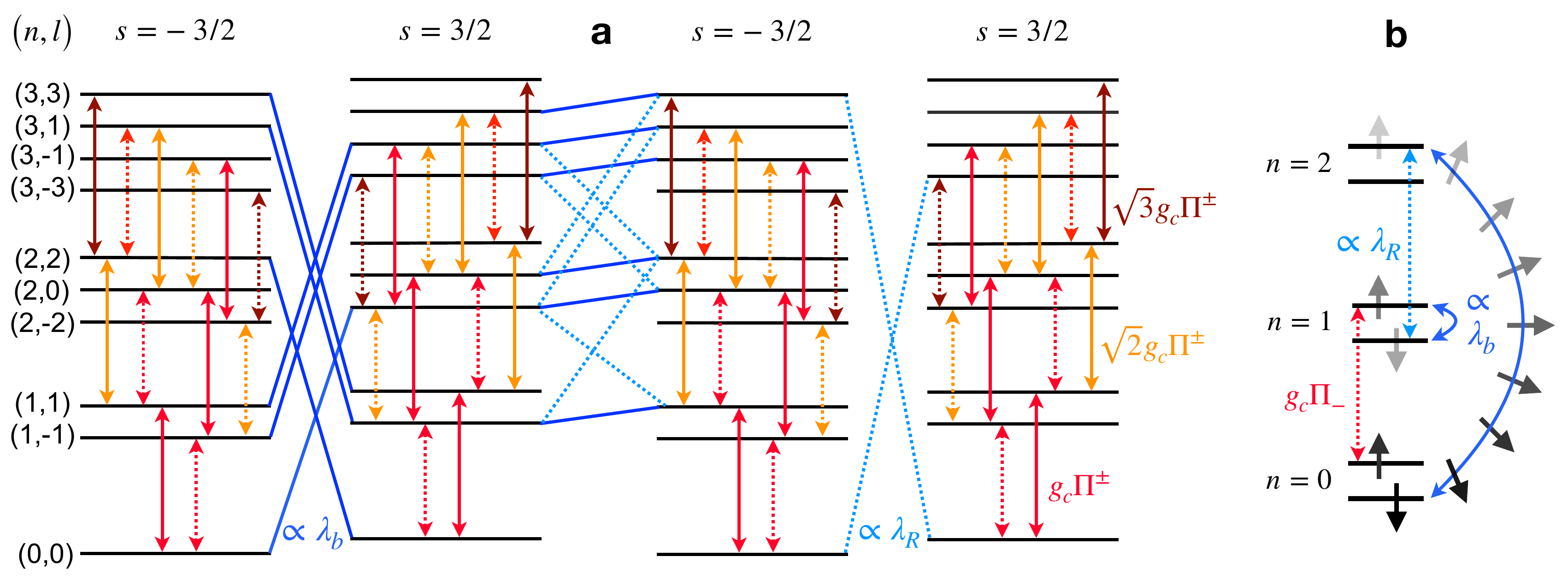}
			\caption{Coupling of the low-energy states. \textbf{a} Allowed transitions between the 20 lowest energy levels (each level shown twice for clarity). In each spin block ($s =  \pm 3/2$), we have dipole transitions, represented by arrows drawn solid ($\Pi^+ = (\omega + \omega_L)/ \omega_0 $) or dashed ($\Pi^- = (\omega - \omega_L)/ \omega_0$) and coloured according to the magnitude of the transition matrix element. The Rashba SOI (dashed blue lines) and in-plane magnetic field (solid blue lines) mix the two spin blocks  with coupling strengths $\lambda_R$ and $\lambda_b$, respectively. \textbf{b} A sketch of the coupling mechanism. A spin in the orbital ground state ($n=0$) can be flipped by the combined effects of the SOI, the in-plane magnetic field and the cavity field. For the sake of clarity we do not show the azimuthal sublevels.}
			\label{fig:single_QD_coupling_schematic}
	\end{figure*}
To support our qualitative reasoning mathematically, we consider the first four orbital levels, $n \in \lbrace 0,1,2,3 \rbrace$. Including spin and azimuthal sublevels, this amounts to a $20 \times 20$ Hamiltonian matrix. In this state space we perform a Schrieffer-Wolff transformation on the total QD Hamiltonian $H_{\text{QD}} = H_0 + H_{\text{SO}} + H_{Z}^x$ to decouple the $n=0$ subspace from higher energy states to quartic order in the perturbation parameters $\lambda_R/\hbar \omega_0$ and $\lambda_b/ \hbar \omega_0$. The SOI mixes 
the eigenstates $|n,l,\pm 3/2\rangle$ of the Hamiltonian $H_0$ yielding the perturbed eigenstates $\vert\overline{ n, l, \pm } \rangle$. We proceed to apply the same transformation to the part describing matter in the interaction Hamiltonian $H_I$. A detailed description of the procedure is given in the supplementary material, Ref.~\footnote{See supplementary material.}.  Projecting the transformed Hamiltonian onto the orbital ground state, we obtain an effective Rabi-type Hamiltonian in the logical basis $ \left\lbrace \vert \uparrow  \; \rangle = \vert \overline{ 0,0,+} \rangle, \vert \downarrow \; \rangle = \vert \overline{ 0,0,-} \rangle \right\rbrace$,
	\begin{align} \label{eq:H_Rabi}
		\tilde{H} =  \frac{\epsilon}{2} \sigma_z  + \hbar \omega_c a^{\dagger}a + \hbar g_s \left(a   + a^{\dagger} \right)  \sigma_y.
	\end{align}
The renormalized energy splitting $\epsilon$ and the effective ground-state spin coupling $g_s$ are given by
	\begin{widetext}
	\begin{subequations}
	\begin{align}
	\label{eq:effective_energy_splitting}
			&\epsilon (B,b) =  \mu_B \sqrt{g_z^2B^2 + g_x^2b^2}  + 2\hbar \sum_{\pm} \frac{ \left( \omega_L \pm \omega \right)^4}{\omega^3 \omega_0} \left[ \left( \frac{\lambda_b}{\hbar \omega_0} \right)^2 \frac{\omega \omega_0}{\omega_Z - 2 \omega_L \mp 2 \omega} + \left( \frac{\lambda_R}{\hbar \omega_0} \right)^2 \frac{3 \left(\omega_L \pm \omega \right)^2}{\omega_Z - 3 \omega_L \pm 3\omega} \right], \\
	 \label{eq:effective_coupling}
		&g_s (B,b) = 24 \frac{\lambda_R}{\hbar \omega_0} \left( \frac{\lambda_b}{\hbar \omega_0} \right)^2 \frac{ \sqrt{\omega_0}  \left(2 \omega_Z  \omega_L + \omega_0^2 \right) \left[ 2 \omega_0^2 \left(  \omega_0^2 + 2\omega_L^2  \right) + \omega_Z \omega_L \left(  3\omega_0^2 + 4\omega_L^2  \right) \right]}{ \omega^{5/2} \left( \omega_Z + \omega - \omega_L \right) \left( \omega_Z + 2\omega - 2\omega_L \right) \left( -\omega_Z + \omega + \omega_L \right) \left(- \omega_Z + 2 \omega + 2 \omega_L \right)}   g_c,
	\end{align}
	\end{subequations}
	\end{widetext}
where we defined the out-of-plane Zeeman energy $\hbar \omega_Z = g_z \mu_B B$. A contour plot of the relative effective coupling strength $g_s/g_c$ as a function of in- and out-of-plane magnetic field components is shown in Fig.~\figref[a]{coupling_contour_plot}. We find excellent agreement between the analytical approximation and numerical results obtained from exact diagonalization of the total Hamiltonian for the energies as a function of both out-of-plane and in-plane magnetic fields (Figs.~\figref[b]{coupling_contour_plot} and \figref[c]{coupling_contour_plot}). Note that, for the sake of clarity,  we display the energy splitting $\epsilon$ only up to quadratic order in the perturbation parameters in Eq.~\eqref{eq:effective_energy_splitting}. As a side result, we mention that linearisation in the magnetic field (valid for $B \lesssim 1$~T) yields an effective out-of-plane g-factor $g_z^{\text{eff}}$, which is reduced not only by the Rashba SOI but also by the in-plane magnetic field (Fig.~\figref[c]{coupling_contour_plot}), an effect that further closes the gap between measured and theoretically predicted values \cite{Hofmann2019arXiv}.  Regarding the effective spin coupling, we do not display the contributions stemming from the in-plane Zeeman energy. We find the ratio $g_s (g_x \neq 0 )/ g_s (g_x = 0)$ to deviate by less than five percent from unity for all magnetic field values of interest except near a resonance at $B^* = \hbar \omega_0/\mu_B \sqrt{g_z^2+ 2g_zm_0/m}$. This divergence is non-physical as it describes the point where the unperturbed eigenstates $\vert 0,0, +3/2\rangle$ and $\vert 1,-1,-3/2 \rangle$ align in energy. The in-plane Zeeman term $H_{Z}^x$ in combination with the dipole transitions induced by the cavity couples these states, and the perturbative approach is not valid in this region (shown as a box in Fig.~\figref[a]{coupling_contour_plot}).
\begin{figure}
		\includegraphics[scale=0.42]{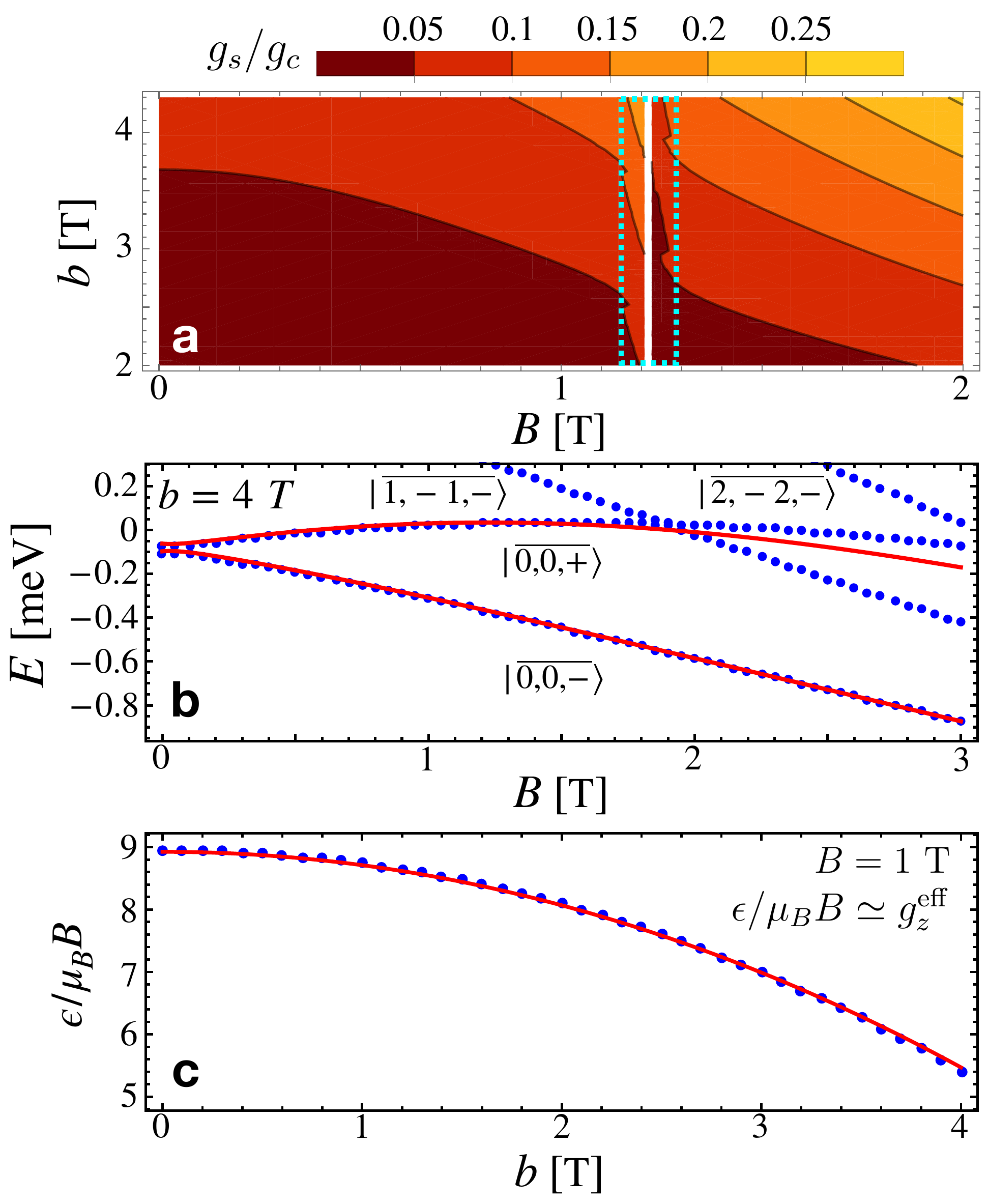}
		\caption{Effective low-energy system. \textbf{a}~The effective spin-photon coupling constant $g_s$ as a function of the in-plane ($b$) and out-of-plane ($B$) magnetic fields. The dotted rectangle indicates that the perturbative approach is not valid in this region (see text).  \textbf{b}, \textbf{c} ~Comparison between the analytical approximation (red solid line) and the numerical result obtained by exact diagonalization of the total Hamiltonian, taking into account the 20 lowest energy states (blue dots). \textbf{b}~Lowest perturbed energy states $\vert \overline{ n,l, \pm } \rangle$  as a function of the out-of-plane magnetic field $B$ at $b=4$~T. We find good agreement between approximation and numerics up to  $B \simeq 2$~T, where the perturbative approach becomes inaccurate due to increased mixing with higher orbital states. The zero-field splitting is caused by the in-plane Zeeman energy and the Rashba SOI. \textbf{c}~Renormalized ground state spin energy separation at $B = 1$~T as a function of the in-plane magnetic field $b$. The g-factor renormalization at $b = 0$~T is due to the Rashba SOI. The parameter values used for all plots are $\gamma_s = 5.11$, $\kappa = 3.41$, $g_z = 10$, $g_x = 0.2$, $\Delta = 10$~meV, $m = 0.2m_0$, $\hbar \omega_0 = 1$meV and $\hbar \alpha_R \langle E_z \rangle =  10^{-11}$~eVm.}
		\label{fig:coupling_contour_plot}
	\end{figure}
	
The effective spin coupling $g_s$ exhibits an exceptionally strong dependence on the QD height $d_z$ via the HH-LH splitting $\Delta$, $g_s \propto 1/\Delta^3 \propto d_z^6$.  An increase in the QD height by a factor of three may thus increase the effective spin coupling strength by two to three orders of magnitude. However, in the present model this restricts the in-plane magnetic field to small values such that the perturbative approach is valid and orbital effects may be neglected. The regime of large in-plane magnetic fields and a relatively large QD height may show particularly strong spin-photon couplings and presents a promising avenue for future research. On the other hand, we find a inverse dependence of the coupling strength on the dot radius, stemming from the behaviour of the momentum matrix elements in $H_{\text{SO}}$. For the same reason, we see a strong dependence on the effective hole mass, with increased coupling for larger masses. The in-plane HH mass in Ge is measured to be $m = 0.09m_0$~\cite{Hendrickx2020} and is extrapolated to be even lower at low hole densities. However, it has been reported that it is possible to increase the HH mass in inversion-symmetric materials by applying tensile strain to the system \cite{Sawano2009,Logan2009}. Finally, $g_s$ depends linearly on the strength of the cubic Rashba SOI. The Rashba coefficient in planar Ge is reported to be $\hbar \alpha_R \langle E_z \rangle = 10^{-13} -10^{-11}$~eVm~\cite{Moriya2014, Morrison2014, Morrison2016}. In Ge nanowires values reach $\hbar \alpha_R \langle E_z \rangle = 10^{-10}$~eVm by applying external gates~\cite{Hao2010}. Such schemes have also been proposed for planar QDs in group III-V materials such as GaAs \cite{deSousa2003}, suggesting the possibility of increasing the Rashba SOI artificially in planar group-IV QDs. Ideally, quantum engineering should therefore aim for relatively large ratios of QD height to QD radius (while maintaining the confinement along z strongest), sizeable HH masses and a large Rashba coefficient. One can see from Fig.~\figref[a]{coupling_contour_plot} that an optimized set of parameters as described above allows for spin-photon couplings exceeding $g_s \approx g_c/4$.
Another way in which one might attempt to artificially increase the spin-photon coupling would be the application of a static electric field in the QD plane. However, for harmonic confinement this merely leads to a shift of the QD position, rather than additional transitions that could modify the cavity coupling.

Finally, we may estimate feasible spin rotation times. Superconducting resonators have typical coupling strengths of $G/2\pi = g_c \omega_c/ 2\pi \omega_0 \simeq 1-10$~MHz for a QD of lateral size $\sqrt{\hbar/m\omega_0} \simeq 10-100$~nm, while cavity loss rates are around $\kappa_c \simeq1$~MHz (yielding a quality factor $Q = f_c/\kappa_c = 10^3$ for typical resonator frequencies of $f_c = 5-10$~GHz)~\cite{Burkard2020}. One major advantage of spin qubits especially in hole systems compared to charge qubits is their high robustness, reaching coherence times $\tau =\Gamma^{-1}$ of several microseconds and even milliseconds \cite{Kolodrubetz2009, Hu2012, Higginbotham2014,Lawrie2020arXiv}. In optimal operation mode, defined by the optimization of parameters described in the previous paragraph, we may reach spin-photon coupling strengths $g_s/2\pi \approx g_c/8\pi$, thus  entering the strong-coupling regime, $g_s > \kappa_c, \Gamma$, allowing for coherent spin rotations with vacuum Rabi frequencies $f_R = g_s / \pi$ in the MHz range. As an application, two-qubit gates may be implemented by harnessing the spin-photon coupling to obtain a controlled interaction between distant spins. For instance, the iSWAP gate may be performed in the dispersive regime in time $\tau = (4k+1)\pi \vert \delta \vert /2g_s^2$ with spin qubit-cavity detuning $\delta = \omega_c - \epsilon/\hbar$ and $k =0,1,2,\ldots$~\cite{Benito2019a}. Consequently, HH systems in Ge allow for fast two-qubit logic with typical gate operation times $\tau \sim \vert \delta \vert $~[kHz]~ns.
	
In conclusion, we show that the pseudo-spin of HHs in planar QDs in bulk inversion-symmetric materials can be coupled to a cavity. The qubit manipulation requires a tilted magnetic field with respect to the QD plane and utilizes the intrinsic cubic Rashba SOI, which is sizeable in Ge. We find that within our model in-plane Zeeman energies and in-plane static electric fields do not considerably enlarge the spin-photon coupling strength. Finally, we propose an optimal planar QD design which allows for coherent Rabi oscillations in the MHz range in the strong coupling regime  and thereby fast long-distance two-qubit logic. Our results consolidate HH spin qubits in Ge as prime candidates for a platform in quantum information processing.

\bibliographystyle{apsrev4-2}
\bibliography{Ge_holes_literature}
\end{document}